\documentclass[manuscript]{aastex}
\usepackage{multirow}

\shorttitle{Multiplicity of Population II field stars}
\shortauthors{D.A. Rastegaev}

\begin{document}

\title{Multiplicity and period distribution of Population II field stars in solar vicinity}
\author{D. A. Rastegaev}
\affil{Special Astrophysical Observatory RAS, Nizhnij Arkhyz, 369167, Russia}
\email{leda@sao.ru}

\begin{abstract}
We examine a sample of 223 F, G and early K metal-poor subdwarfs
$(\mathrm{[m/H]}<-1)$ with high proper motions \mbox{($\mu >
0.2''/$ year)} at the distances of up to $250$ pc from the Sun.
By means of our own speckle interferometric observations conducted
on the 6 m BTA telescope of the Special Astrophysical Observatory of
the Russian Academy of Sciences, and the
spectroscopic and visual data taken from the literature, we
determine the frequency of binary and multiple
systems in this sample. The ratio of single, binary, triple and
quadruple systems among 221 primary components of the sample is
147:64:9:1. We show that the distribution of orbital periods of
binary and multiple subdwarfs is asymmetric in the range of up to
$P=10^{10}$ days, and has a maximum at $P=10^{2}-10^{3}$ days,
what differs from the distribution, obtained for the thin disc G
dwarfs \citep{dm91}. We estimated the number of
undetected companions in our sample. Comparing the frequency of binary
subdwarfs in the field and in the globular clusters, we show that
the process of halo field star formation by the means of
destruction of globular clusters is very unlikely in our Galaxy.
We discuss the multiplicity of old metal-poor stars in nearby
stellar streams.
\end{abstract}



\section{Introduction}

The study of metal-poor stars $(\mathrm{[Fe/H]}<-1)$ makes
it possible to shed light on numerous problems of modern
astrophysics, such as heavy elements production in supernova
explosions, the metallicity distribution function of the stellar
halo, the initial mass function, the nature of the Big Bang and
the first generation, or Population III, stars, etc. An important
place among these fundamental problems is occupied by the
questions of the origin, chemical and dynamical evolution of our
Galaxy. The oldest stars with the masses of
$M\leqslant0.8\ M_{\odot}$ are unevolved. Therefore, the abundance of
chemical elements in their atmospheres reproduces the composition
of prestellar matter. Additional information on spatial motions of
these stars preserves the possibility to reconstruct the way the
Milky Way formed.

The orbital elements of binary and multiple stellar systems are
an important tool for studying prestellar matter. In single low
mass stars, the mass is the only parameter conserved since the
time of star formation. Binary and multiple systems bear three
more conserved values: the angular momentum, the eccentricity and
the mass ratio of their components \citep{larson} in case of
detached systems. Therefore,
binary and multiple stars carry more information on the process
of star formation than single stars. The study of binary and
multiple metal-poor systems enables us to impose certain
restrictions on the physical conditions in prestellar matter at
the time of the genesis of our Galaxy. Metal-poor stars are
common in the globular clusters, galactic halo and in the
galactic field, where an existence of the so-called stellar
streams was revealed (e.g., \citealt{eggen_1996a,eggen_1996b}). The
multiplicity and the orbital parameters of binary and multiple
stars in these streams may also provide additional information on
the nature of the stream's progenitor and its dynamical evolution.

The problem of stellar multiplicity was widely discussed in the
literature, however, it mostly concerned the thin disc
stars with solar-like metallicities \citep{dm91,fischer_marcy,halbwachs}.
Metal-poor stars were studied much less, as their occurrence in the solar
neighbourhood is less than 1\%, according to \citet{nordstrom} catalog.
Early works, addressing the rate of binary systems among Population II
stars, showed that this value is small, as compared to the
analogous value for Population I stars \citep{abt_levi,
crampton_hartwick,abt_willmarth}. In subsequent works
\citep{preston_sneden,goldberg_2002,latham_2002} it
was concluded that these values are
indistinguishable (idem \citealt{abt}). A long-term spectroscopic
monitoring of about $1\ 500$ nearby stars with high proper
motions (\citealt{clla} (hereinafter CLLA),
\citeyear{carney_2001}; \citealt{goldberg_2002,latham_2002})
has played an important role in the study of the
multiplicity of metal-poor stars. Spectroscopic studies cover the
systems with relatively short orbital periods ($P\lesssim10$
years).

The study of long-period couples with common proper motion
components \citep{zapatero} confirms the
hypothesis of an equal frequency of binary systems among the
old and young stellar populations (idem \citealt*{allen}).
Meanwhile, an `intermediate' period range of $P\approx10-1\ 000$
years, which corresponds to the semi-major orbital axes of
$a\approx10-100$ AU in the solar neighbourhood, remains
poorly understood to date. This range can
be studied with the use of adaptive optics, speckle
interferometry and long baseline interferometry. The scarce
Population II stars observations, made by means of the
interferometric techniques, were ran either for the brightest
stars \citep{lu} or with relatively low angular resolution
\citep*{zinnecker}. Notwithstanding the empirical data
available to date, the number of known binary and multiple
systems with metal-poor components remains small.

In order to enlarge the database of binary and multiple Population
II stars, to define their orbital parameters and the properties of
their components, we conducted speckle interferometric
observations of 223 metal-poor subdwarfs with high proper motions
located in the solar neighbourhood \citep*{rastegaev_2007,rastegaev_2008}.
The observations were made with the diffraction-limited
resolution of the 6 m Big Telescope Alt-azimuthal
(BTA) of the Special Astrophysical Observatory of the Russian
Academy of Sciences ($0.023 ''$ at the wavelength
of 550 nm). In the present work, we analyse the multiplicity and orbital
periods distribution for binary and multiple stars.
We made our analysis based on own observations
and the data adopted from other authors.
Additionally, an attempt was made to examine the ratio of binary
and multiple stars in the streams of old metal-poor stars located
in the solar neighbourhood.

\section{Sample}

For the observations with high angular resolution, we compiled a
sample of 223 field subdwarfs of the F, G and early K spectral
classes \citep{rastegaev_2007} from the CLLA
catalog. The CLLA presents a
spectroscopically studied sample of the A--K spectral types dwarfs
from the \textit{Lowell Proper Motion Survey} \citep*{lpms_1971,lpms_1978},
which mainly includes the stars from the Northern
Hemisphere with proper motions exceeding $0.26''$ per annum and
brighter than $16^{m}$ in the $B$ band.

We selected 223 stars from the CLLA using the following criteria:
\begin{itemize}
\item
metallicity $\mathrm{[m/H]}<-1$,
\item
declination $\delta>-10^\circ$,
\item
apparent magnitude $\mathrm{m_V}<12^{m}$.
\end{itemize}
The last criterion was determined by the limiting stellar
magnitude of our speckle interferometer \citep{maximov},
which was about $13^{m}$. No restrictions were applied on the
heliocentric distances of these stars, evenly distributed on the
celestial sphere. The maximum distance to the sample objects is
250 pc. The median heliocentric distance of the selected stars is
approximately $100$ pc. This allows us to take advantage of high
angular resolution in order to detect new systems, since at such
distances the pairs with semi-major orbital axes from 10 to 100
AU are hard to detect both spectroscopically and visually.

Using the two following criteria, $\mu>0.26\ ''/$year and
$\mathrm{[m/H]}<-1$, we tried to rule out the thick and thin disc
stars. In case of the thick disc though, these limits are not
stringent and some objects may belong to the metal-weak tail of
the thick disc \citep{arifyanto}.

To separate the halo stars from the disc stars in our sample, we used a formal method,
described in the appendix of \citet{grether_lineweaver_2007}.
The equations establishing the probability that a star
belongs to the thin disc ($P_{\rm thin}$), the thick disc ($P_{\rm thick}$) or the halo
($P_{\rm halo}$) are
\begin{equation}
\label{population_eq}
P_{\rm thin} = f_1 \frac{P_1}{P} \; , \; P_{\rm thick} = f_2 \frac{P_2}{P} \; , \; P_{\rm halo} = f_3 \frac{P_3}{P},
\end{equation}
where
\begin{displaymath}
P  = \sum f_{i} P_{i},
\end{displaymath}
\begin{eqnarray}
P_{i} = C_{i} \exp \left[ -\frac{U^2}{2\sigma_{U_{i}}^2}
-\frac{(V- \langle V\rangle)^2}{2\sigma_{V_{i}}^2}-\frac{W^2}{2\sigma_{W_{i}}^2}- \right. \nonumber \\
\left.  -\frac{([\mathrm{Fe/H}]- \langle [\mathrm{Fe/H}]_{i}\rangle)^2}{2\sigma_{[\mathrm{Fe/H}]_{i}}^2} \right], \nonumber
\end{eqnarray}
\begin{displaymath}
C_{i} = \frac{1}{\sigma_{U_{i}} \sigma_{V_{i}} \sigma_{W_{i}} \sigma_{[\mathrm{Fe/H}]_{i}}},
\end{displaymath}
\begin{displaymath}
i = \mathrm{1\ (thin\ disc),\ 2\ (thick\ disc),\ 3\ (halo).}
\end{displaymath}
The  input parameters for these equations are adopted from \citet{robin_2003}, and presented
in Table \ref{populations_properties}.  We rejected three out of 221 systems:
two double stars G99-48 and G166-45, and one single BD $-1^{\circ} 1792$, because for these
systems there are no heliocentric distances or $UVW$ components in the CLLA.
We referred the objects with $P_{\rm halo} > 0.5$
to the halo stars. There is a total of 148 of such stars in our sample. The remaining 70 objects
belong to the thick disc. None of the stars from the sample have $P_{\rm thin}$ exceeding $0.01$.
Average metallicity and space velocity vector components for halo stars in our sample are:
$(\langle [\mathrm{Fe/H}]_{halo}\rangle, \langle U_{halo}\rangle,
\langle V_{halo}\rangle, \langle W_{halo}\rangle)=$ ($-1.9\pm0.5$, $-16\pm152$ km/s,
$-180\pm81$ km/s, $-1\pm78$ km/s), where the errors are standard deviations.
In a similar way, these values for the thick disc stars are:
$(\langle [\mathrm{Fe/H}]_{thick}\rangle, \langle U_{thick}\rangle,
\langle V_{thick}\rangle, \langle W_{thick}\rangle)=$ ($-1.2\pm0.2$, $-12\pm83$ km/s, $-92\pm59$
km/s, $3\pm50$ km/s). However, the system of equations (\ref{population_eq}) might be unable to
accurately describe the situation in the areas where the star
parameters of different Populations overlap.

\begin{figure}
\includegraphics[width=180mm,height=120mm]{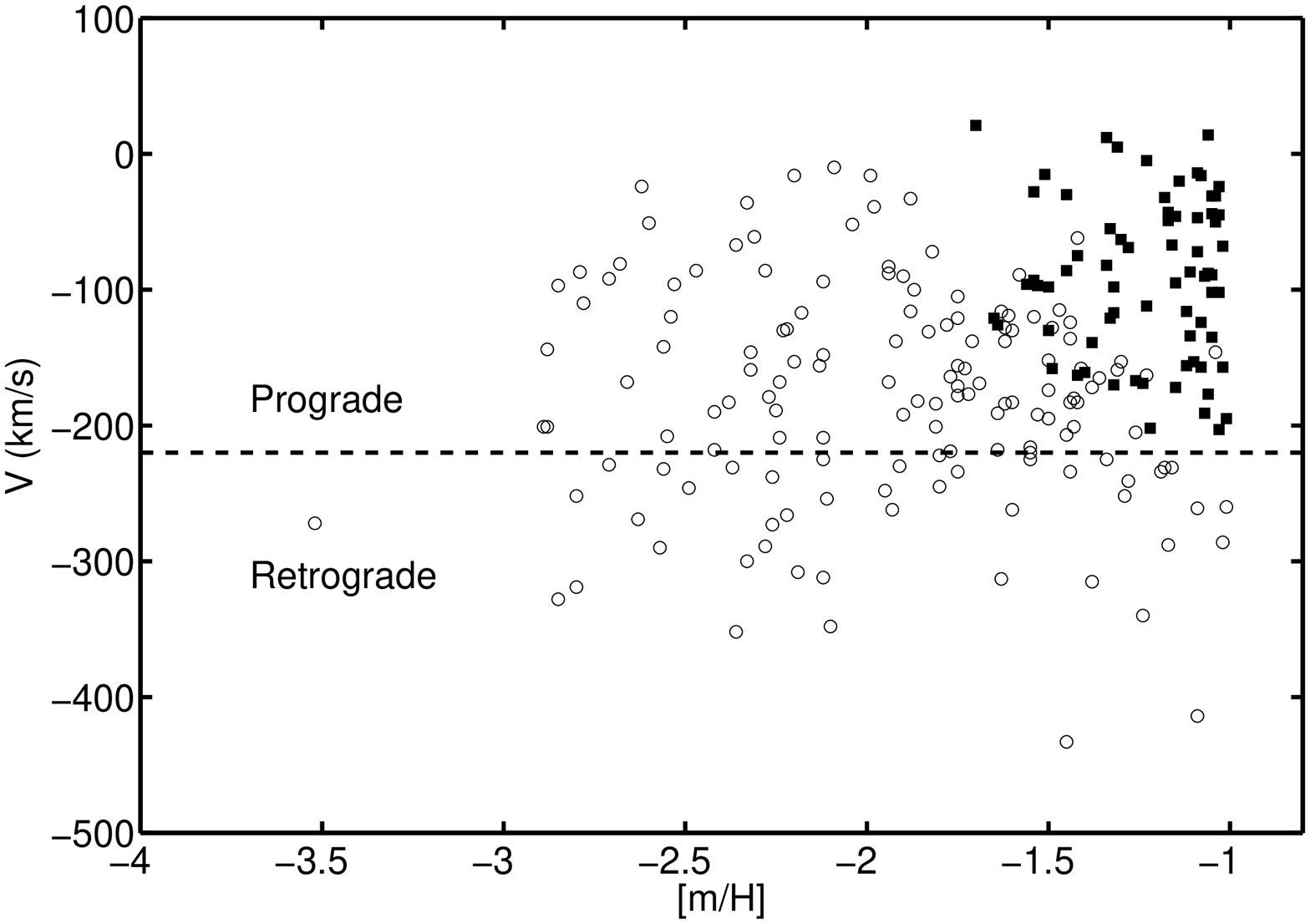}
\caption{Metallicity [m/H] versus the $V$-component of spatial
velocity for the stars in our sample. The circles represent
the halo stars and the squares mark the thick disc stars.
The dashed line separates the stars on prograde (upper half)
and retrograde (lower half) orbits.} \label{vm}
\end{figure}

Fig. \ref{vm} shows
the distribution of the sample stars on the graph metallicity versus
$V$-component of spatial velocity, data taken from the CLLA. The
halo stars with a substantial dispersion of spatial velocities
and with low metallicities are located in the central and left
parts of the figure. At the top right of the figure, along with
the halo stars, there are stars of the thick disc's metal-weak
tail \citep{arifyanto}.

\begin{figure}
\includegraphics[width=180mm,height=120mm]{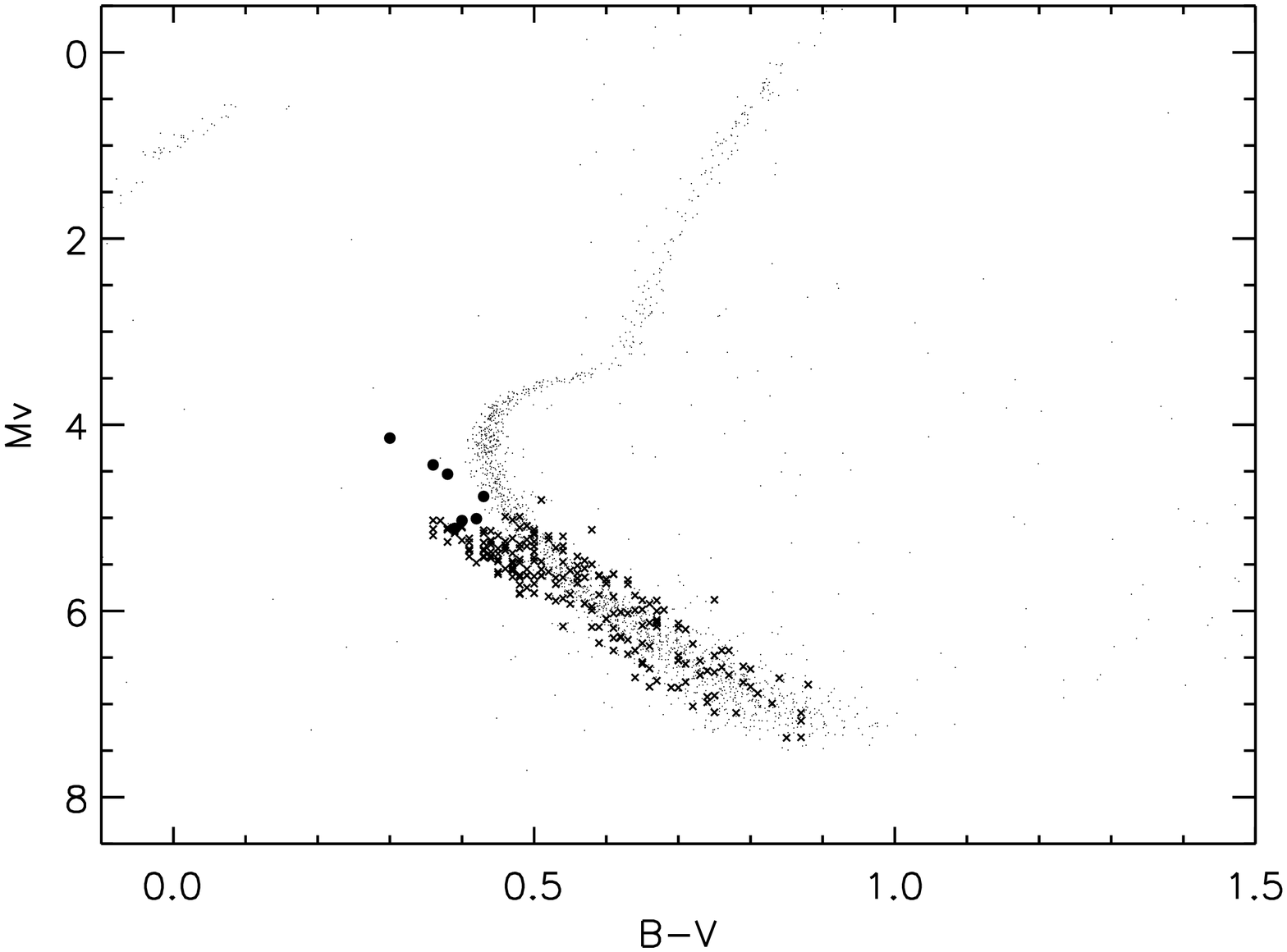}
\caption{
$(B-V)$ -- $M_{V}$ diagram for 213 stars in our sample
(crosses). To make a comparison, we plotted the stars from the
M13 globular cluster (dots) from \citet{rey_2001}. The distance
modulus for M13 is $(m-M)_0=14.38\pm0.10$ \citep{grundahl}.
Seven blue stragglers from our sample are marked with
filled circles.} \label{cmd}
\end{figure}

The sample consists of main sequence stars and seven blue
stragglers \citep*{carney_2001,carney_2005b}, --- a
continuation of the main sequence in the area of hotter and bluer
stars, as compared to the turnoff stars. Fig. \ref{cmd}
represents the location of the sample stars in the
Hertzsprung-Russell diagram compared to the old population of the
M13 globular cluster with $\mathrm{[Fe/H]}=-1.61$ \citep*{grundahl}.

\begin{figure}
\includegraphics[width=180mm,height=120mm]{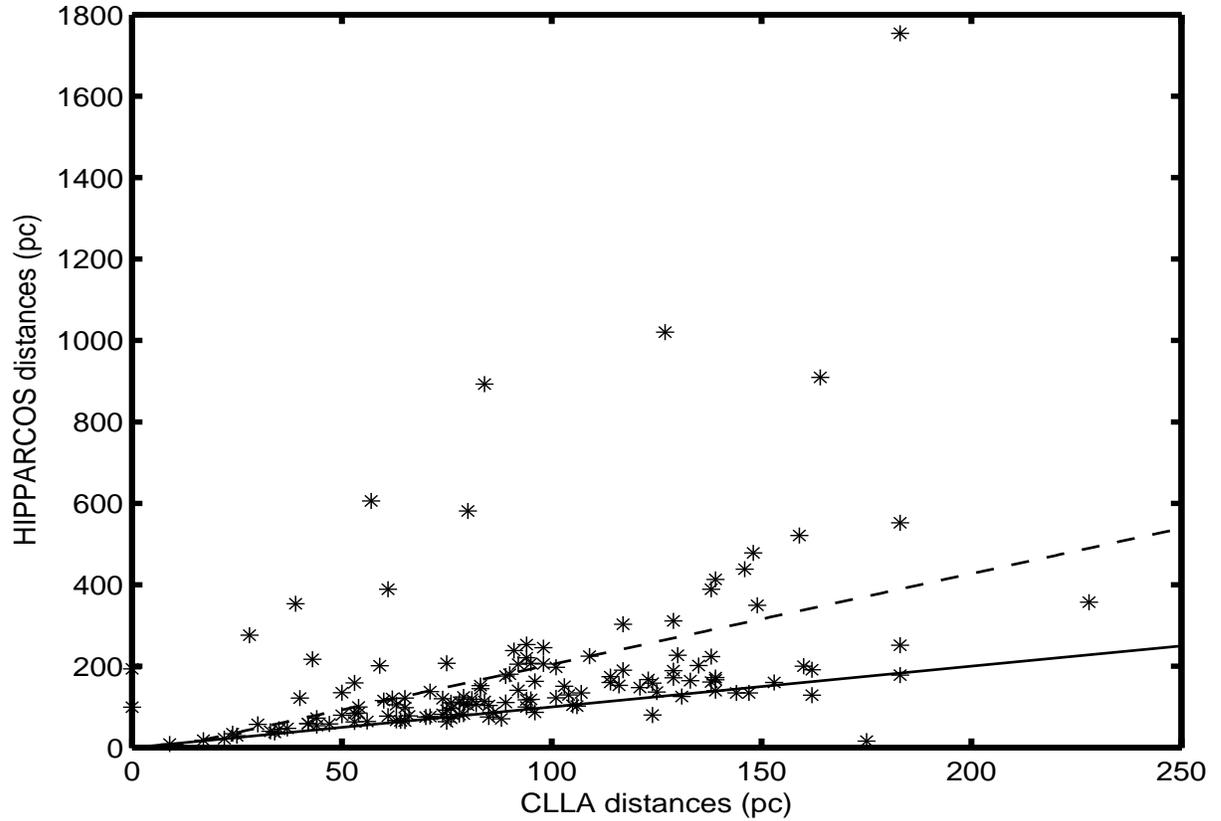}
\caption{A comparison between the trigonometric distances from the
{\it HIPPARCOS} catalog \citep{leeuwen_2007} and photometric distances
from the CLLA for 133 sample stars. The dotted line is a linear regression
$y=-18.58+2.23x$. $r=0.42$ is the correlation coefficient. The solid
line represents the $y=x$ function.} \label{dist}
\end{figure}

Fig. \ref{dist} represents a comparison between the
trigonometric and photometric distances for the sample stars.
The parallaxes for 135 sample stars were taken from the {\it
HIPPARCOS} satellite data \citep{leeuwen_2007}. The corresponding
photometric distances for 133 stars are retrieved from the CLLA.
We can see that the distances, obtained with {\it HIPPARCOS}, are
systematically larger than the corresponding distances from the
CLLA. This difference can be explained by some unaccounted
components, the presence of which may lead to an underestimation
of photometric distances. Another reason for this discrepancy is
the photometric distance calibration adopted in the CLLA.

Most of the stars in our sample were examined for common proper
motion components \citep{allen,zapatero}.

\section[]{Observations and results}

The speckle interferometric observations of 223 sample stars were
carried out in 2006--2007 on the 6 m BTA telescope
\citep{rastegaev_2007,rastegaev_2008} which diffraction-limited
resolution is $0.023 ''$ for $\lambda=550$ nm and $0.033 ''$
for $\lambda=800$ nm. Most of the
observations were carried out using the system \citep{maksimov}
based on a 512$\times$512 EMCCD (a CCD featuring on-chip
multiplication gain) with high quantum efficiency and linearity.
This system allowed us to detect objects with magnitude
differences between the components of up to $\triangle m = 5^{m}$.
Taking into account the limiting stellar magnitude of our sample
($\mathrm{m_V}<12^{m}$), detected secondary component can be as faint
as $17^{m}$. The $4.4''$ field of view of our system allows detection
of secondary components at a separation of $3''$ from the primary
star. The speckle interferograms were recorded using five
filters: $545/30$, $550/20$, $600/40$, $800/100$ and $800/110$ nm
(the first number indicates the central wavelength of the filter,
the second --- the half-width of the filter's bandwidth) with the
exposures of 5 to 20 ms. For each object, we accumulated from 500
to $2\ 000$ short exposure images depending on weather
conditions. The observations were made with an average seeing of
$1.5''$. The accuracy of our speckle interferogram processing
method \citep{balega_2002} may be as good as $0.02^m$,
$0.001''$, and $0.1^{\circ}$ for the component magnitude
difference, angular separation and position angle respectively.

For 19 stars in our sample we observed the speckle
interferometric companions. Sixteen companions were resolved
astrometrically for the first time. We discovered 5 new binary
systems (\object{G191-55}, \object{G114-25}, \object{G142-44},
\object{G28-43}, \object{G130-7}), 3 triple systems (\object{G87-47},
\object{G111-38}, \object{G190-10}) and one quadruple system,
\object{G89-14}. The position parameters and magnitude differences
between the speckle interferometric components are listed in tables
1 and 2 of \citet{rastegaev_2008}.

\section{Sample completeness}

To be able to determine the completeness of our sample, we used
the star count method. From all known subdwarfs within 25 pc from
the Sun, we selected 5 objects with absolute magnitudes
$M_{V}<8^m$ and metallicities $\mathrm{[Fe/H]}<-1$ (see tab. 1 in
\citealt{fuchs_jahreiss}). The star GJ 1064 A, with the
metallicity of $-1$ dex, was as well included in our sample. An
extrapolation to the volume of our sample (with the radius of
$\approx250$ pc) increases the number of such objects to
$\approx5 \ 000$. Obviously, when we consider the frequency of
stars in different volumes, we cannot depart from the uniform
distribution of stars in space. We have to bear in mind the
structure of the Galaxy and the data available on the star
distribution in its various subsystems. However, as we show in
the Appendix, for the stars in our sample located within 250 pc from
the Sun, the structure of our Galaxy can be neglected. Therefore,
the number of objects in our sample constitutes about $5\%$ from
the total number of stars in question located in the examined
region of space. We have to mention that while examining the
stars in different volumes, we were taking account of the primary
components only in case of spectroscopic and speckle interferometric
pairs, and of both components in the case of visual and common
proper motion pairs.

\section{Undetected companions}
\label{und_com_section}

Capability to detect a binary system is determined by both the
observations method used and by the physical characteristics of
the system itself. As the sample objects were examined using
three different methods, --- spectroscopic, speckle interferometric and
visual, we have to analyse the number of systems unaccounted for
by each of these methods. To do that, we conditionally divided the
spectroscopic and astrometric pairs by the values of their semi-major
axis at $a<10$ AU and $a>$ 10 AU, and their orbital periods at
$P<10\ 000 $ days and $P>10\ 000$ days, respectively.

For spectroscopic pairs, we hypothesized that the component mass ratio
is uniformly distributed from 0 to 1, then took the deduced period
distribution for the studied stars in the range of 0 to $10\ 000$
days (see below), and made our calculations using the formulae
from \citet*{mazeh}. Our conclusion is that the probability of
not detecting a binary system approximately equals to $20\%$.

As for the visual and interferometric binaries, the secondary can
be detected if the following two conditions are satisfied.
Firstly, the magnitude difference between the components should
not exceed a certain critical value, determined by the detector's
dynamical range and the spectral range used. Secondly, in
case of speckle interferometry, the angular separation between the
components should be larger than the telescope's diffraction limit
and smaller than the detector's field of view, or, in
case of visual studies, the angular separation should be smaller than
the certain restrictions imposed.

In order to estimate the number of unresolved systems due to the
effect of ellipse projection of the true orbit on the picture
plane, and due to the secondary component's orbital phase at the
moment of observation, we used the Monte Carlo simulation. We
modelled a sample of stellar orbits, evenly distributed in space,
with heliocentric distances from 25 to 250 pc, uniformly
distributed eccentricities (from 0 to 0.9) and arguments of periapsis
(from 0 to 360 $^\circ$). Inclinations of the true orbits
to the picture plane obey the \mbox{sin $i$} law.

The resulting orbits were projected onto the picture plane using
the following formulae \citep{couteau}
\begin{eqnarray}
\nonumber
\rho=\frac{a\ (1-e^2)}{1+e\ \mathrm{cos}\ \upsilon}\ \frac{\mathrm{cos}\ (\upsilon+\omega)}{\mathrm{cos}\ (\theta-\Omega)},\\
\mathrm{tg}\ (\theta-\Omega) = \mathrm{tg} \ (\upsilon+\omega)\ \mathrm{cos}\ i,\\ \nonumber
\mathrm{tg} \ \frac{\upsilon}{2}=\sqrt{\frac{1+e}{1-e}}\ \mathrm{tg} \ \frac{u}{2},\\ \nonumber
M=u-e\ \mathrm{sin}\ u,
\end{eqnarray}
where $\rho$ is the angular separation between the components in
the picture plane, $a$ is the orbital semi-major axis (in arcseconds),
$e$ is eccentricity, $\upsilon$ is the true anomaly, $u$
is the eccentric anomaly, $M$ is the mean anomaly, $\omega$ is the
argument of periapsis, $\theta$ is the position angle of a
secondary component, $\Omega$ is the longitude of an ascending
node and $i$ is the inclination angle between orbital and picture
planes.

Further, a random orbital position of the satellite was
predetermined for each simulated system.
At this point we accounted for the fact that a companion
spends more time at apoastron than at periastron.
Then the projection distances $\rho$
from the satellite to the primary star were counted. A relative
number of projection distances greater than $0.033''$(the
diffraction limit of the 6 m telescope in a 800/100 filter) is,
in fact, the probability value of a system detection using the
speckle interferometric and visual methods. The results of
modelling show that for the systems with an orbital semi-major
axis exceeding $a=10$ AU, such probability is close to 1 and an
influence of the components' geometry can be neglected. This
probability weakly depends on the choice of a plausible
distribution of semi-major axis for $a>10$ AU.

\begin{figure}
\includegraphics[width=180mm,height=120mm]{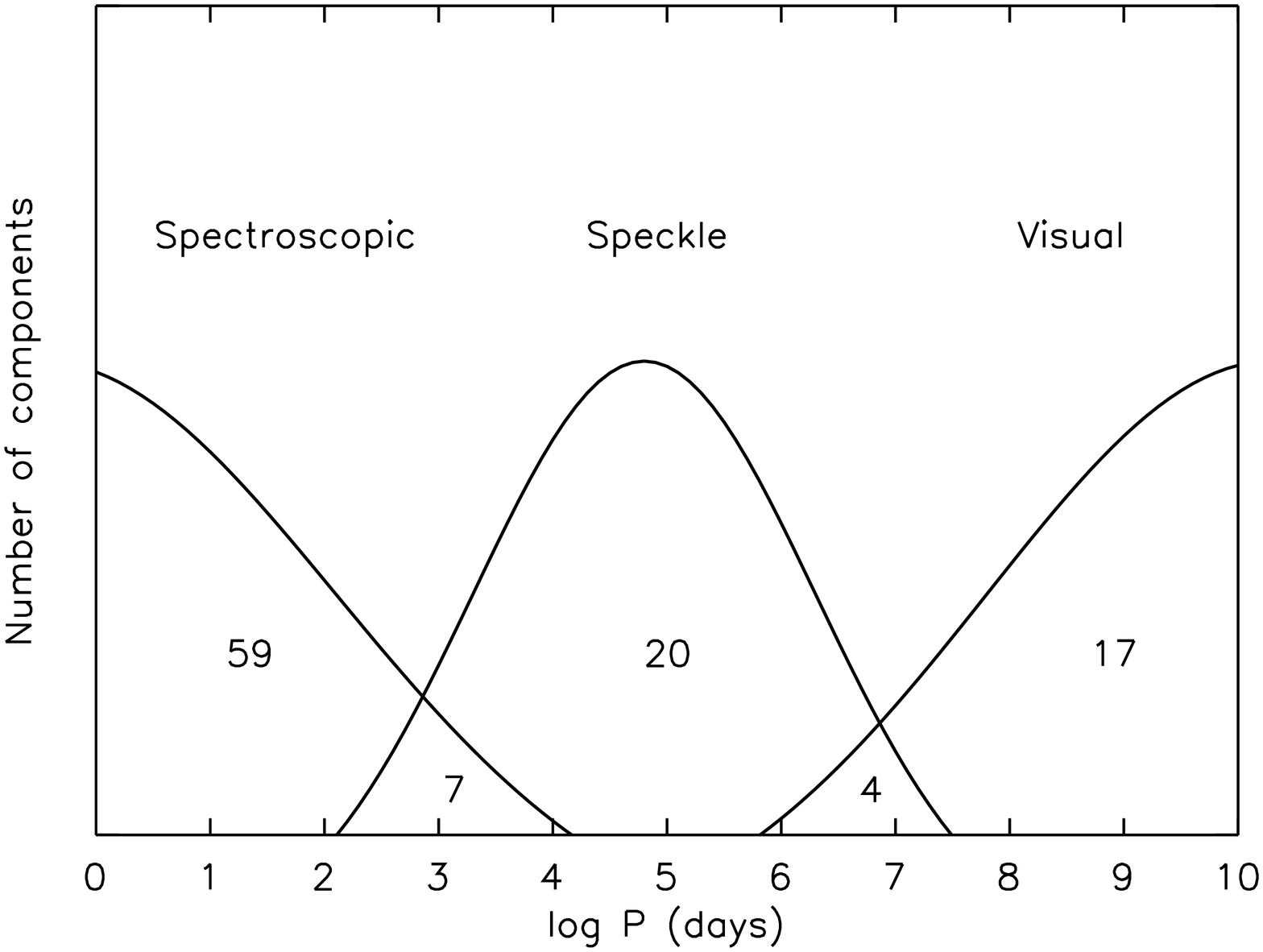}
\caption{Schematic representation of the companions in our sample
detected using different methods.} \label{diff_meth_sys}
\end{figure}

Let's consider now the incompleteness of detection caused by the
magnitude difference between the components. An average mass of a
primary star in our sample is $0.65\ M_{\odot}$ and a standard
deviation is $\sigma_{{\textrm{\tiny M}}}\approx0.05\ M_{\odot}$.
The maximum magnitude difference between the components, detected
by speckle interferometric and visual methods, is about $5^m$
\citep{maksimov,zapatero}. For
speckle interferometry, we used a more conservative estimate of
$4^{m}$. According to the model of \citet{baraffe} for
$\mathrm{[Fe/H]}=-1$, the minimal mass of a secondary, which is
4 mag fainter than a primary of an average mass, equals
$M_{min}=0.20\ M_{\odot}$ in the $I$ band. This band roughly
corresponds to our 800/100 and 800/110 filters.
To compute the number of undetected companions, we have to know
the function of mass ratio distribution. We used
a uniform distribution. If the mass ratio for wide (speckle
interferometric, visual and CPM) pairs obeys a
uniform $q$ distribution, then from the following ratio
\begin{equation}
\label{integral_q_eq}
\frac{\int\limits^{M_2=M_{min}}_{M_2=0.08}\ d M_2}
{\int\limits^{M_2=0.8}_{M_2=M_{min}} \ d M_2},
\end{equation}
we find the quantity of secondary components with masses in the
range from $0.08\ M_{\odot}$ (mass of a brown dwarf) to $M_{min}$
per one secondary component with the mass in the range from
$M_{min}$ to $0.8\ M_{\odot}$ (a maximal mass of the stars in
our sample). This quantity equals 0.2. Therefore,
a fifth of discovered speckle interferometric companions
remains undetected at the given distribution of $f(q)$ with
the use of 800/100 (or 800/110) filter. For the
550/20 and 545/30 filters, $M_{min}=0.28\ M_{\odot}$ and the
expression (\ref{integral_q_eq}) is approximately 0.38, which makes these
filters less suitable for our task in the sense of magnitude
difference (yet, more suitable from the viewpoint of the angular
resolution). The search for common proper motion pairs for the
sample stars was conducted in the $I$ band \citep{zapatero}.
Forty seven objects from our sample were observed
speckle interferometrically, solely using the 550/20 or 545/30
filters, 9 objects were observed only in the 600/40 filter and one
object (G183-9) was observed in the 550/20 and 600/40 filters.
The 166 remaining stars were at least once observed in the
800/100 or 800/110 filters. With this in mind, we find that the
expression (\ref{integral_q_eq}) is approximately equal to 0.25 for
all filters. Therefore, according to these rough estimates, for 33
astrometric components in our sample (see Fig. \ref{diff_meth_sys})
we have about 8 unaccounted companions or 24\%; that is
comparable to 20\% of unaccounted spectral companions. Thus, we
can assume that the number of unaccounted components in our
sample does not depend on the orbital period and is $20-25\%$.

Analogously, the IMF-like distribution $f(q)\sim q^{-1.3}$ for
$0.08 \leq M/M_{\odot} < 0.5$ \citep{kroupa_2001} gives us 37
unaccounted companions. In case $f(q)$ grows towards bigger $q$
values \citep{soderhjelm_2007}, then the number of undetected astrometric
components in our sample does not exceed 5. For our further estimates
we chose the values corresponding to the uniform $q$ distribution.

\section{Multiplicity of the sample}

\subsection{Raw estimates}
In order to calculate the ratio of the systems of different
multiplicity among the studied stars, we complemented the results
of our speckle interferometric measurements by the data from
spectroscopic and visual studies found in literature. The information
on the spectroscopic companions was adopted from the publications
dedicated to long-term spectroscopic monitoring of Population II stars
by Carney, Latham, Laird et al. (CLLA; \citealt{carney_2001,goldberg_2002,
latham_2002,latham_pc}). The data on wide visual pairs was taken
from \citet{allen} and \citet{zapatero} via
the WDS catalog \citep{mason}. As a result, the ratio of
single:binary:triple:quadruple ($S:B:T:Q$) systems for the stars in
our sample is 147:64:9:1 \citep{rastegaev_2007,rastegaev_2008}. Therefore,
at least 159 stars from 306 stars in our sample (221 main components
and 85 satellites) belong to binary and multiple systems. The
multiplicity of our sample is\footnote{Everywhere further,
we use a 95\% confidence interval as an error for the multiplicity obtained
by us. For this purpose we use the properties of the binomial distribution.}
$33\%^{+7\%}_{-6\%}$, where multiplicity is
understood as the ratio of binary and multiple systems to the
total number of systems
\begin{displaymath}
f_{systems}=\frac{B+T+Q}{S+B+T+Q}.
\end{displaymath}

A similar, unadjusted for unresolved companions value for the
thin disc stars of spectral classes from F7 to G9,
equals to 51:40:7:2 \citep{dm91}, with the multiplicity of about
50\%. It is necessary to pay attention to the differences between
the two compared samples. When our sample was formed from the
stars with a certain apparent magnitude limit and high proper
motions, the sample from Duquennoy and Mayor is only limited by
the heliocentric distances (all their stars are located within 22
pc from the Sun).

In terms of multiplicity, the Population II stars differ from the
Population I. At least a third of metal-poor systems and at least
half of the systems with solar like abundances are binary and
multiple. This discrepancy can be explained by both the
complexity of detection of low mass metal-poor spectroscopic
satellites, by the selection effects, and by the dynamical
evolution of binary and multiple stars.

\subsection{Corrected estimates}

To estimate the true multiplicity of the sample stars, we have to account
for various selection effects that are unavoidable in astronomical observations.
The five underlaying criteria for the choice of stars in our sample are:
proper motion ($\mu > 0.26\ ''/$year), metallicity ($\mathrm{[m/H]}<-1$),
magnitude ($\mathrm{m_V}<12^{m}$), spectral classes (F, G and early K),
and position in the sky (all our stars
are located in the Northern hemisphere). We also have to take into account
the undetected companions (Section \ref{und_com_section}).

For the sake of simplicity we assume that the multiplicity of a stellar
Population does not depend on proper motion and position in the sky.
This gives us grounds to disregard the selection effects related to the proper
motion of the stars and their coordinates.

To account for the bias incurred by magnitude, we have to consider
the $\mathrm{\ddot{O}}$pik effect \citep*{opik,goldberg_2003},
which applies to binaries detected within a magnitude-limited sample
of stars. This effect is an expansion of the Malmquist effect for binary
stars. Binaries are on the average brighter than singles. Thus, in a
magnitude-limited sample the binaries are observed from a bigger volume
in space than the single stars. To avoid this effect we rejected the
binaries which are brighter than $12^m$ due to the contribution of the
secondary component to the total luminosity.
For SB1 binaries we adopted that the luminosity of the secondary
component is not less than 2 magnitudes fainter than the primary
\citep{goldberg_2002}. We discarded the SB1 pairs
whose primary component was fainter than $12^m$ in the assumption
that the secondary component is 2 magnitudes fainter than the primary.
For SB2 pairs the luminosity of the primary component was determined from the total
luminosity of the system and the mass ratio of the components \citep{goldberg_2002}
using evolutionary tracks from \cite{baraffe}. We thus rejected four binaries,
one SB1 (G242-14) and 3 SB2 (G86-40, G99-48, G183-9) as the measured
luminosities of the primaries are fainter than 12 magnitude in the $V$-band.
In the worst case for SB2 systems we have to discard 4 pairs in the assumption that
for a given integrated magnitude the luminosities of primary and secondary components
are equal. None of the speckle interferometric pairs have been excluded from consideration
due to the $\mathrm{\ddot{O}}$pik effect. The magnitude differences between the
components measured by us for a given integrated magnitude suggest that the primary
components of the speckle pairs are brighter than $12^m$.
Not a single system of higher multiplicity was dropped from the
analysis as their primaries are not compliant with the condition $m_V > 12^m$.

Taking into account an adjustment for unresolved components
and $\mathrm{\ddot{O}}$pik effect, the multiplicity of F, G
and early K subdwarfs in the solar neighbourhood is, according to
our calculations, at least $40\%$ and at least $60\%$ for the thin
disc G dwarfs \citep{dm91}. In both cases correction
does not exceed $10\%$.

\subsection{Multiplicity as a function of metallicity}

\begin{figure}
\includegraphics[width=180mm,height=120mm]{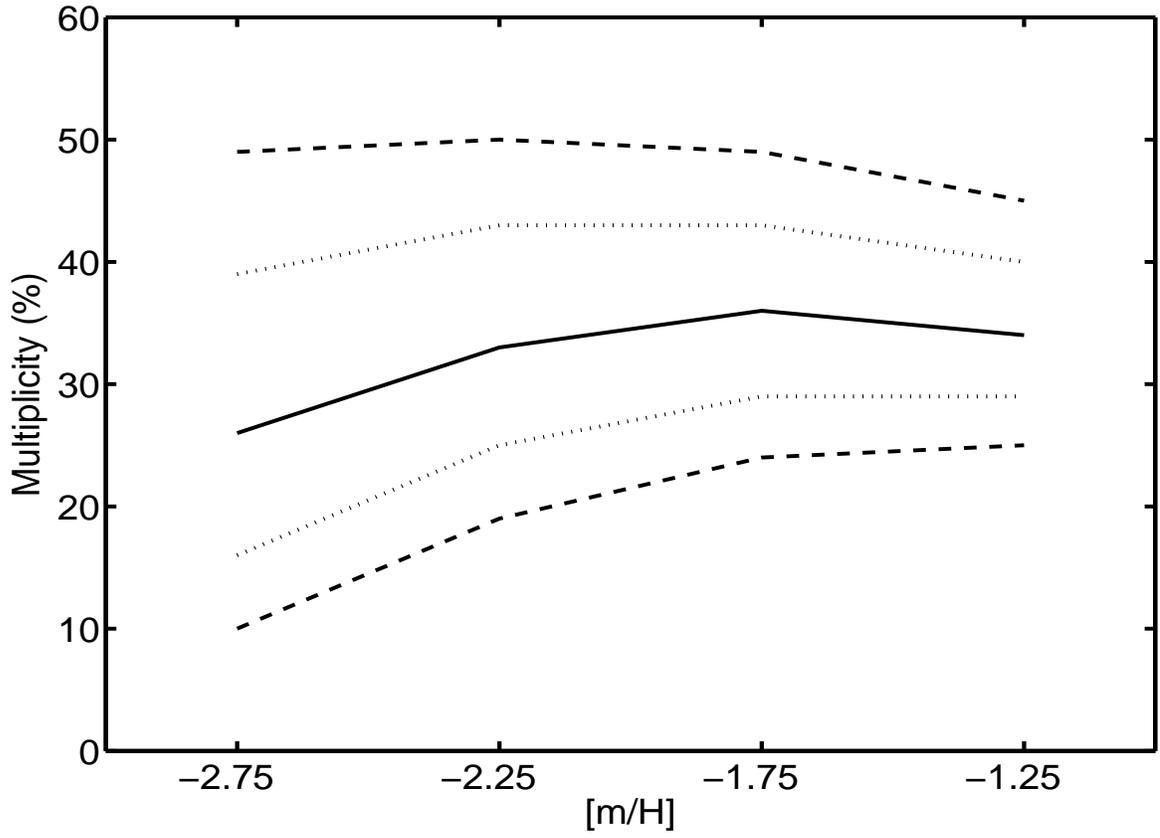}
\caption{Multiplicity versus metallicity for stars in our sample
(solid line). Dashed lines mark the 95 \% confidence interval.
Dotted lines mark the 70 \% confidence interval.
} \label{m_H_mult}
\end{figure}

To investigate the effect of metallicity on the multiplicity of stars in our
sample, we divided it into four metallicity bins: $(-3,-2.5]$, $(-2.5,-2.0]$,
$(-2.0,-1.5]$, $(-1.5,-1.0)$. One single star, G64-12 with $\mathrm{[m/H]}=-3.52$
was rejected from the analysis. In every bin we evaluated the ratio of
single:binary:triple:quadruple systems and multiplicity $f_{systems}$.
The acquired results are presented in Table \ref{mult_metal_tab}, whereof
it is clear that in the range $\mathrm{[m/H]}$ from $-2.5$ to $-1.0$ the
multiplicity does not depend on metallicity and constitutes one third.
In the range $(-3,-2.5]$ the multiplicity is somewhat lower, but the number of
systems in this range is smaller than in the rest of ranges, which affects the
accuracy of determination of $f_{systems}$ (see Fig. \ref{m_H_mult}).
It can be seen from Fig. \ref{m_H_mult} that in the first approximation
the rate of binary and multiple
systems in our sample feebly depends on metallicity. \citet{carney_2005a} obtained
analogous results based on a bigger sample of stars, studied spectroscopically.
Note that the bulk of our triple stars belong to the range $\mathrm{[m/H]}=(-1.5,-1.0)$.

\subsection{Halo versus thick disc}

Using the set of equations (\ref{population_eq}) we separated the stars of the thick disc
and the halo stars in our sample. The ratio single:binary:triple:quadruple systems for
thick disc stars is 46:19:5:0, and the multiplicity $f_{systems}=34\%^{+13\%}_{-11\%}$.
The same for the halo stars $S:B:T:Q=$101:45:4:1, and $f_{systems}=33\%^{+8\%}_{-7\%}$.

An independence of the ratio of binary and multiple systems from the conditional division
'thick disc-halo', as well as the consistency of $f_{systems} \approx 33\%$ under metallicity
changes (Table \ref{mult_metal_tab}) might testify that most of the stars in our sample belong
to one and the same galactic subsystem, the halo. Note again that the set of equations
(\ref{population_eq}) might not be operable in cases when the stellar parameter ranges of
different galactic subsystems overlap. It is not impossible though that the Population II stars,
both the halo and thick disc stars have similar ratios of the systems of different multiplicity.

\subsection{Multiplicity as a function of kinematics}

Table \ref{mult_kinematics_tab} represents the basic facts on the
dependence of multiplicity of the sample stars on kinematics.
We excluded three objects from consideration: two double
stars G99-48 and G166-45, and one single BD $-1^{\circ} 1792$,
because for them there are no distances or $UVW$
components in the CLLA. For
single and binary and mltiple stars the average values and standard
deviations of the space velocity
vector components are: $\langle U_{s}\rangle=-13\pm137$ km/s, $\langle
U_{b+m}\rangle=-18\pm127$ km/s, $\langle V_{s}\rangle=-160\pm86$ km/s, $\langle
V_{b+m}\rangle=-135\pm81$ km/s, $\langle W_{s}\rangle=2\pm74$ km/s, $\langle
W_{b+m}\rangle=-3\pm61$ km/s. For norm of velocity vector
$v=\sqrt{U^2+V^2+W^2}$: $\langle v_{s}\rangle=222\pm89$
km/s, $\langle v_{b+m}\rangle=196\pm80$ km/s. In the first approximation we may
consider that for our stars the multiplicity is constant and
does not depend on the spatial velocity vector components.

\begin{figure}
\includegraphics[width=180mm,height=120mm]{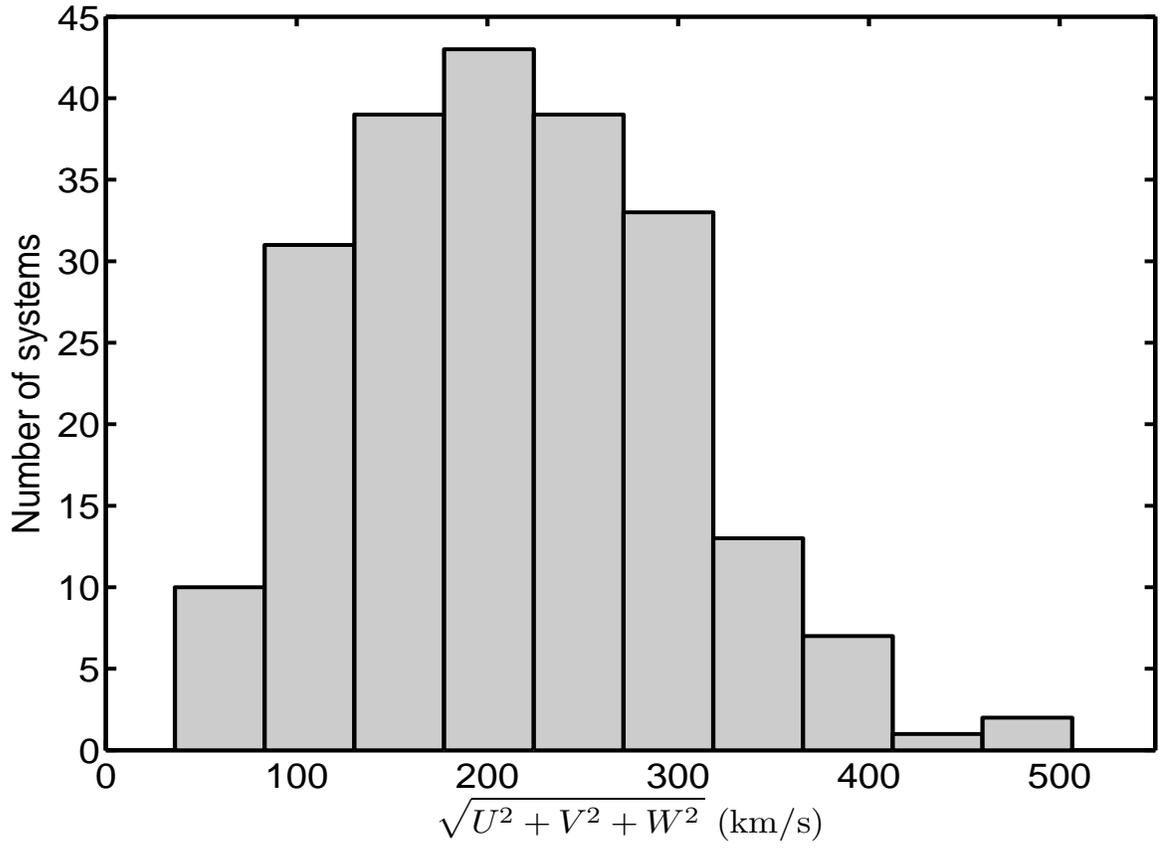}
\caption{Norm of velocity vector distribution for stars in our sample.} \label{norm_UVW}
\end{figure}

\begin{figure*}
\includegraphics[width=180mm,height=158mm]{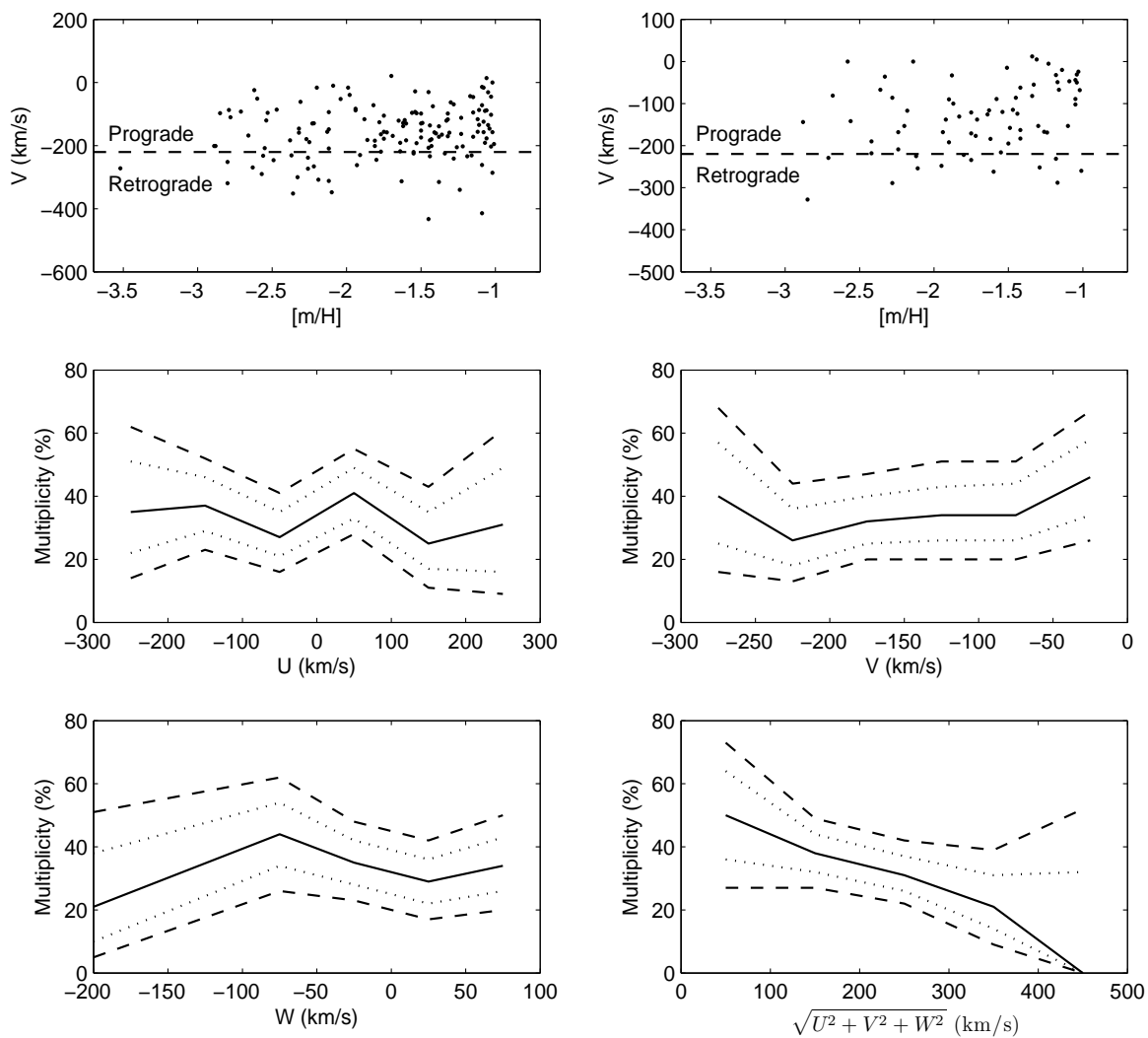}
\caption{
The upper panel shows $V$ versus $\mathrm{[m/H]}$
dependence for single (left part) and double and multiple (right part)
stars. Middle and lower panels show the frequency of double and multiple stars
depending on the components of the velocity vector and its norm. Dashed
lines mark the 95\% confidence interval. Dotted lines mark the 70\% confidence interval.
}
\label{kinematic_fig}
\end{figure*}

The distribution of the norm of spatial velocity vector $v=\sqrt{U^2+V^2+W^2}$ for stars in
our sample is shown on Fig. \ref{norm_UVW}. The average value of this norm is 214 km/s
with the standard deviation 86 km/s. Fig. \ref{kinematic_fig} shows the $V$ versus $\mathrm{[m/H]}$
dependence for single (left upper panel) and double and multiple (right upper panel) stars in
our sample. Middle and lower panels of this figure show the dependence of the multiplicity of
our stars on four kinematic parameters $UVW$ and $v$. The figure shows that the frequency of
double and multiple stars in our sample weakly depends on the change of $UVW$ components. We can
draw lines corresponding to a constant multiplicity of approximately 35\% within a 70\% confidence
level (see middle panel and left lower panel). For norm of spatial velocity vector a trend of
decreasing multiplicity with increasing $v$ is seen (see right lower panel). The data in the
95\% confidence level do not contradict the hypothesis $f_{systems}(v)=const$ or a small growth of
multiplicity with the increase of $v$. Nevertheless, with the probability of at least 70\% we can
say that with the increase of the norm of spatial velocity vector the frequency of double and multiple
systems in our sample falls. Most likely, the bigger the spatial velocity of a star, the less the
probability that it has a companion.

Forty five from 223 stars in our sample are moving on
retrograde galactic orbits ($V<-220$ km/s), i.e. in the opposite
direction to the rotation of our Galaxy. For these stars, the
ratio of single:binary:triple:quadruple systems is 32:12:1:0 and
their multiplicity is $29\%^{+15\%}_{-13\%}$. \citet{carney_2005a},
analyzing a sample of 374 stars on highly retrograde orbits ($V<-300$ km/s)
showed, that the frequency of spectroscopic binaries among them is
two times smaller than that for the stars moving along with the
Galaxy's rotation. Our data, bearing an analysis of a wide range
of periods, do not contradict their findings, yet, an
insignificant number of highly retrograde objects in our sample
does not allow us to make any definitive conclusions.

\section{Period distribution}

\begin{figure}
\includegraphics[width=180mm,height=120mm]{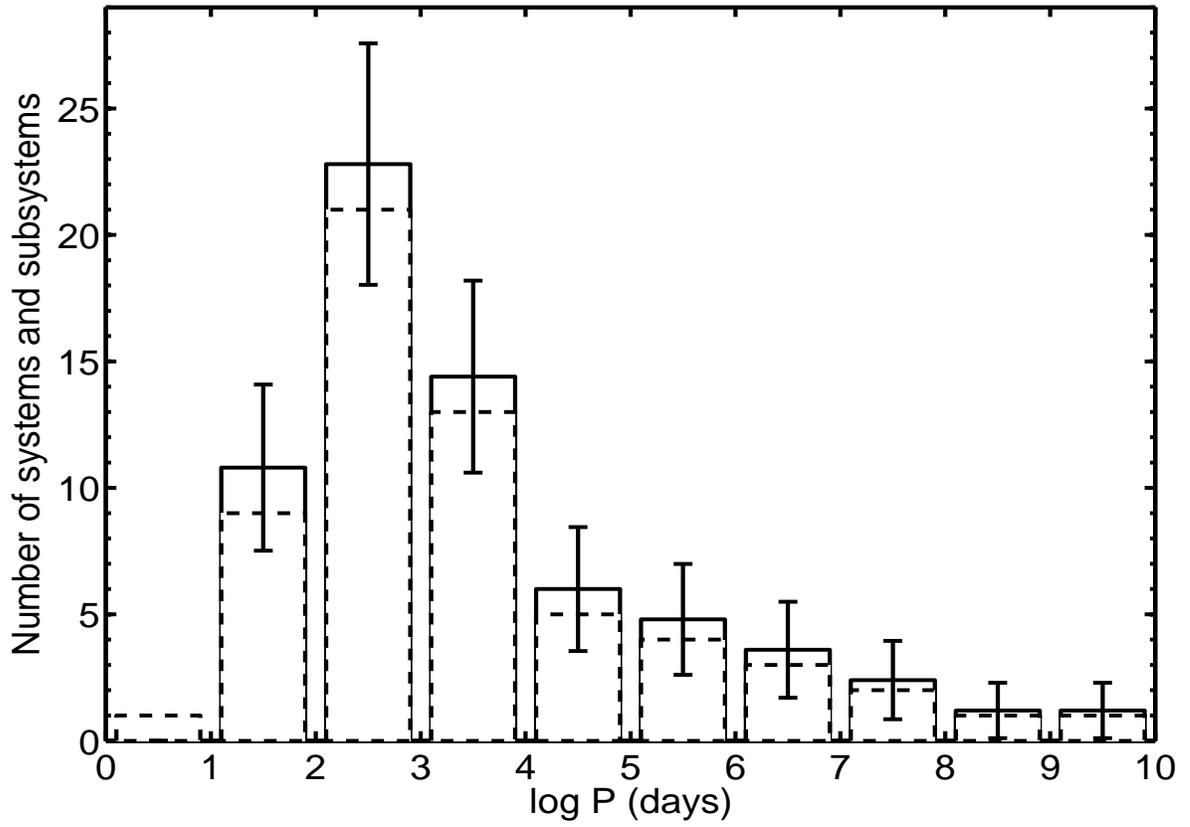}
\caption{Period distribution for 60 binary stars in our sample
(dashed line). The solid line marks the same distribution but corrected
for $\mathrm{\ddot{O}}$pik effect and unresolved components.
Error bars represent square roots from the number of stars in
each bin.} \label{periodsbin_pic}
\end{figure}

\begin{figure}
\includegraphics[width=180mm,height=120mm]{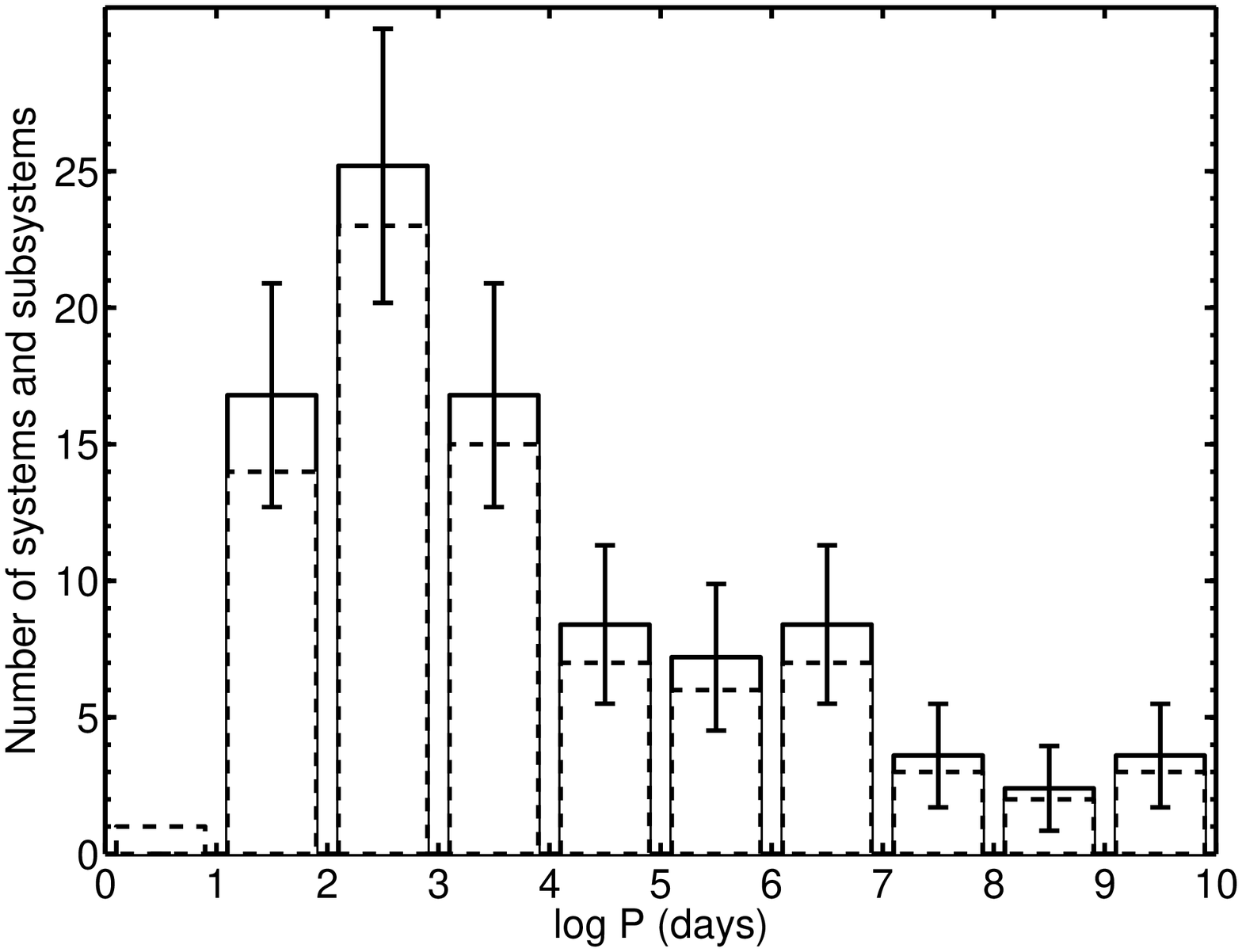}
\caption{Period distribution for 60 binary stars and 21
subsystems of 10 multiple stars in our sample
(dashed line). The solid line marks the same distribution but corrected
for $\mathrm{\ddot{O}}$pik effect and unresolved components.
Error bars represent square roots from the number of stars in each bin.}
\label{periodsall_pic}
\end{figure}

The distribution of orbital periods for binary stars in our
sample is shown in Fig. \ref{periodsbin_pic}. The periods of
spectroscopic pairs are taken from \citet{goldberg_2002} and
\citet{latham_2002}. The periods of astrometric pairs were derived with
the help of the generalized Kepler's third law on the basis of an
empirical relation of the projected angular separation between the
components and the semi-major axis. Knowing the system's parallax
$\pi$ and the projected angular separation between the components
$\rho$, the expected value of the semi-major axis is calculated
using the formula from \citet{allen}
\begin{equation}
\label{allen_eq}
\left\langle a \right\rangle =\mathrm{antilog}\left[ \mathrm{log}\frac{\rho}{\pi} + 0.146 \right] ,
\end{equation}
where $\left\langle a \right\rangle$ is expressed in astronomical
units, $\rho$ and $\pi$ in arcseconds. Taking each system
individually, we derived the sum mass of the components from the
temperatures of the primary (CLLA) and secondary components, and
from the magnitude difference, if the temperature of the secondary
was unknown. To do this, we used models from \citet{baraffe}.
Angular distances $\rho$ between the
components of astrometric pairs were obtained from speckle
interferometric observations or adopted from \citet{allen}
and \citet{zapatero}. If the systems parallaxes
were known from the {\it HIPPARCOS} catalog with an accuracy of
better than $30\%$, we used them instead of the distances cited in the
CLLA catalog. As a result, we were able to determine the
periods for 60 binary systems out of 64 in our sample. The
periods for the four remaining suspected binary systems, ---
\object{G186-26} and \object{G210-33} and two blue stragglers, \object{BD +25$^{\circ}$ 1981}
and \object{G43-3} \citep{carney_2001}, are too long to be determined
\citep{latham_pc}. The period distribution
for 60 binaries and 10 multiple systems (18 subsystems of 9 triple
stars and 3 subsystems of a quadruple star G89-14), is shown in
Fig. ~\ref{periodsall_pic}. The distributions corrected for
$\mathrm{\ddot{O}}$pik effect and unresolved components
on Fig. \ref{periodsbin_pic} and \ref{periodsall_pic} are
marked by a solid line.

\begin{figure}
\includegraphics[width=180mm,height=120mm]{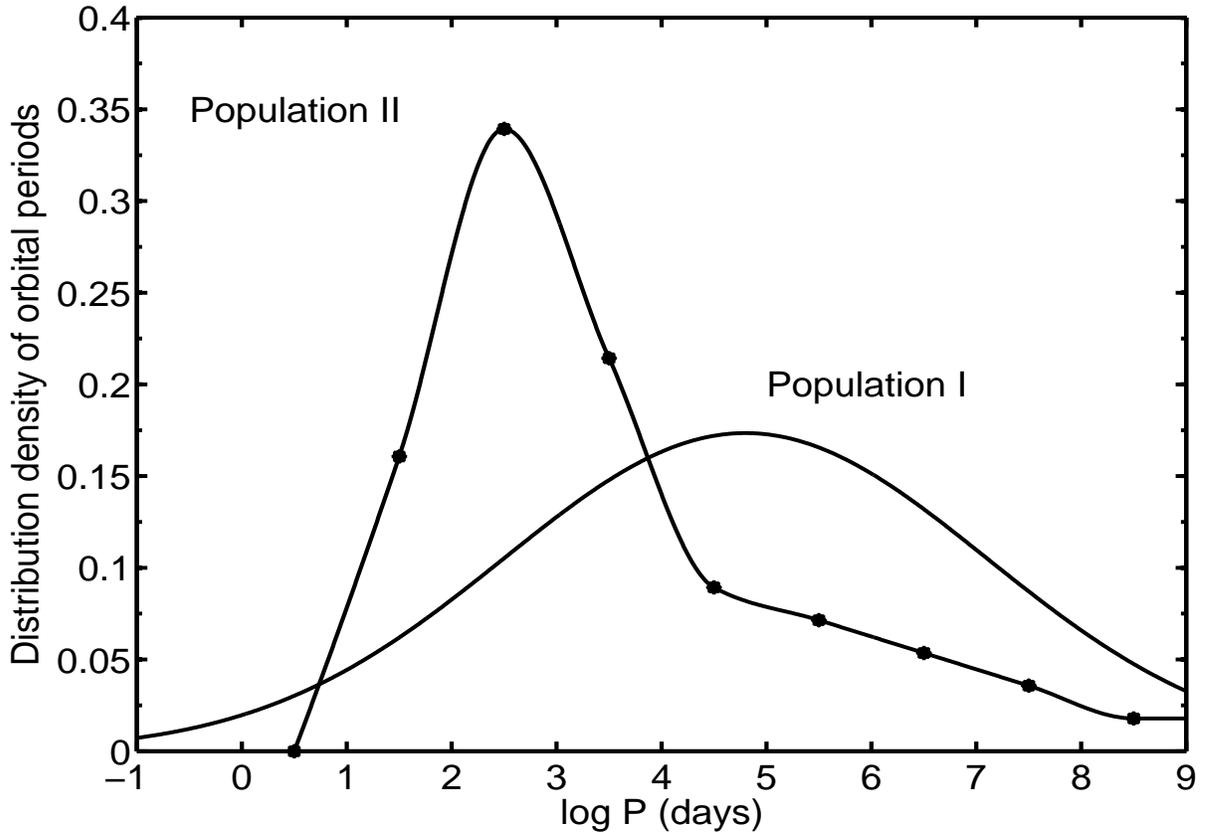}
\caption{Distribution of densities of orbital periods for binary
stars in our sample (Population II) interpolated by a cubic spline
and idem for the thin disc stars (Population I), approximated by a
Gaussian in \citet{dm91}.} \label{periodscomparison_pic}
\end{figure}

Let's compare the resulted distribution with an analogous one
for the thin disc stars (Fig. \ref{periodscomparison_pic}).
The maximum of an unsymmetrical period
distribution for the stars in our sample lays in the range of
$\log P=2-3$ dex (i.e. hundreds of days). For Population I stars,
the $\log P$ distribution, which Duquennoy and Mayor approximated
by a Gaussian, has a maximum in the range of $4-5$ dex (tens of
thousands of days). An important feature of our distribution is a
small number of short period ($\log P < 1$) pairs. This range is
represented by the only SB2 star \object{G183-9} with a period of
about 6 days \citep{goldberg_2002}, which is excluded
while accounting for the $\mathrm{\ddot{O}}$pik effect.

To make a homogeneity check of the two samples, we conducted a
nonparametric $\chi^2$ test \citep{kremer}. By an accidental
coincidence, the number of compared periods equals to 81 for each
population. The test shows that the hypothesis of uniformity
(i.e. common general population of two samples) can be rejected
with a more than 95\% probability.

Both distributions (Fig. \ref{periodscomparison_pic}) may be distorted
by the selection effects. However, the differences in both the shape
of the distributions and the location of the maxima could be induced
by the dynamical evolution, undergone by the Population II stars.
For example, a small quantity of old systems with periods of more
than ten thousand days may be the result of a dynamical evolution
at the stage of Galaxy formation. According to a recent concept, a
bigger part of the stellar halo in our Galaxy was formed from
small galactic systems \citep{bell}. It is quite likely
that at the stage of accretion of small galaxies-satellites of
the Milky Way and their destruction, the physical conditions were
favourable for dissociations of wide pairs with low binding
energy. Attempting to explain the period distribution of
Population II stars by a destructive impact of giant molecular
clouds and other local perturbations of the gravitational
potential of our Galaxy on the old stars orbiting around the
galactic centre \citep*{weinberg}, is quite problematic.
Such objects spend most of their lifespans away from the galactic
plane.

\section{Interesting triple and quadruple systems in our sample}

In this section we will examine two old multiple systems that we
find remarkable: a triple system \object{G40-14} and a quadruple
\object{G89-14}. Such objects are of great interest for the studies
of the dynamical evolution and for the checks of various criteria
of dynamical stability. In Table \ref{multiple}, we listed all the
detected systems from our sample having more than 2 components.
From ten multiple systems, nine are triples and only one is
quadruple.

The uniqueness of the triple system G40-14 is in its retrograde
galactic orbit with $V\approx-230$ km/s (CLLA). The inner
subsystem of G40-14 is an SB1 pair with a period of $60.615$ days
\citep{latham_2002}. The outer subsystem is formed by a visual
component, which is located $98''$ away from the spectroscopic pair.
Assuming that the system's heliocentric distance is 235 pc, the
expected semi-major axis is $\left\langle a\right\rangle \approx
30\ 000$ AU \citep{allen}. We made a check for retrograde
objects in the latest version (dated 13 August 2007) of the
Multiple Star Catalog \citep{tokovinin_1997}, which is a compilation
of 1158 known stellar systems with three or more components. In
order to do that, we calculated the $U$ $V$ $W$ components of
spatial velocities for the catalog stars from the cited
parallaxes, radial velocities and proper motions, using the
formulae from \citet{johnson_soderblom}. It appeared that only one
object, a triple system \object{ADS 16644}, is moving on a retrograde
galactic orbit ($V\approx-330$ km/s). Hence, G40-14 is the second
of all known systems with more than two components moving against
the rotation of the Galaxy.

\begin{figure*}
\includegraphics[width=156mm,height=75mm]{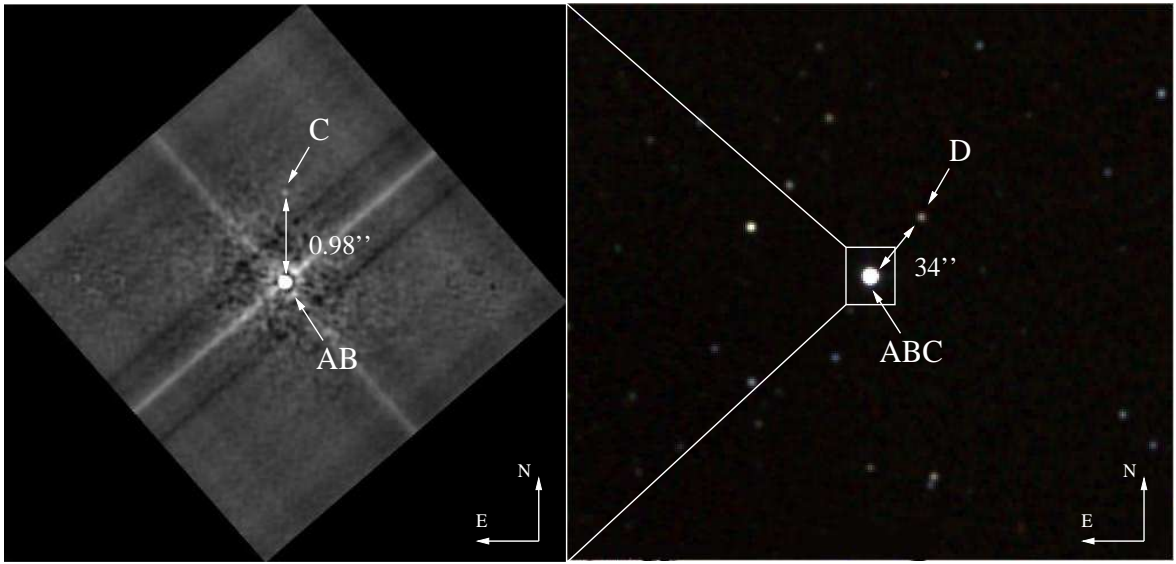}
\caption{On the right: CPM subsystem of G89-14 (POSS archive).
On the left: the speckle interferometric subsystem (image reconstructed
from speckle interferograms using bispectral technique). The magnitude
difference between AB and C is $4.2^m$ in 800/100 filter.} \label{G8914pic}
\end{figure*}

G89-14 (for details, see \citealt{rastegaev_2009}) is a system with the
highest multiplicity in our sample.
It consists of four components (Fig. \ref{G8914pic}): an SB1 pair
AB with a period of 190 days \citep{latham_2002},
a speckle interferometric component C
located at $\approx1''$ from this SB1 pair \citep{rastegaev_2007}
and a common proper motion companion D at $34''$
\citep{allen}. Based on the data from \citet{allen}, the
evolutionary tracks from \citet{baraffe}, and on our speckle
interferometric measurements, we evaluated the period ratio of
the three G89-14 subsystems: 0.52 : 3 000 : 650 000 yr. Another
well-known metal-poor quadruple system is \object{NQ Ser}
\citep{tokovinin_1997} with $\mathrm{[Fe/H]}=-1.05$ \citep{nordstrom}.
This multiple star was repeatedly observed on the
BTA by means of speckle interferometry (e.g., \citealt{balega_2006}).
According to our information, G89-14 with
$\mathrm{[m/H]}=-1.9$ (CLLA) is the most metal-poor quadruple
system known to date, which makes it an interesting object for a
more detailed study.

\section{Multiplicity of metal-poor stellar streams}

Stellar streams (e.g., \citealt{eggen_1996a,eggen_1996b}) are associations
of stars possessing similar kinematics and metallicity. The study of
such streams allows restoring to a certain degree the picture of
the formation of various dynamical structures in our Galaxy.
Traditionally, the stellar streams are being selected in a
certain phase space and then their origin is interpreted using
the data of spectroscopic analysis. In a phase space, a fine structure
like stellar multiplicity can give additional information on the
dynamical evolution of the stream and its primogenitor. However,
until now it was not taken into due consideration.

The following six stars of our sample: \object{G10-4}, \object{G13-9},
\object{G60-48}, \object{G24-3}, \object{G18-54}, \object{G28-43},
are part of the \object{Kapteyn's star} moving group \citep{eggen_1996a},
10 other objects: \object{G130-65}, \object{G75-56}, \object{G5-35},
\object{G40-14}, \object{G114-25}, \object{G11-44}, \object{G13-35},
\object{G183-11}, \object{G182-32}, \object{G126-52}, belong to the
\object{Ross 451} moving group \citep{eggen_1996b}. In Table \ref{streams} we
are listing some characteristics of these two halo streams. In the
penultimate column of the table, you can see the ratio of single,
binary and triple systems for the group members in our sample. In
the last column, an analogous estimate is given for all known
members of the groups, the data taken from literature.
Unfortunately, multiplicity of these stars is poorly studied and
the ratio in the last column can only serve as a lower limit for
the frequency of binary and multiple systems in the streams. The
two moving groups listed above have comparable multiplicities,
both exceeding 10\%.

The fact that we find similar multiplicities in moving groups,
does not contradict the dynamical hypothesis of their origin. In
this case, the stellar streams are formed by a random selection
from the general population of field stars. Recent works on the
problem of origin of stellar streams show that the Hercules
stream \citep{bensby}, as well as the Pleiades, Hyades,
and Sirius moving groups \citep*{famaey_2007,famaey_2008},
formed as a result of dynamical (resonant) influence of
our Galaxy on the field stars. However, the scenario of accreted
stellar streams cannot be ruled out. This requires similar
multiplicities of the progenitors of the flows. Further detailed
studies are required to help answer the question whether there
exist any distinctions between various stellar streams.

\section{Conclusion}

In this paper we examine a sample of 223 subdwarfs belonging to
the F, G and early K spectral classes, located within 250 pc from
the Sun, with the metallicities $\mathrm{[m/H]}<-1$ and proper
motions $\mu \gtrsim 0.2''/$year. Stars make up about $ 5\% $ of
the total number of objects of this type in the studied space
volume. The subdwarfs were observed using the spectroscopic
\citep{goldberg_2002,latham_2002}, interferometric
\citep{rastegaev_2007,rastegaev_2008} and visual methods
\citep{zapatero}. Presented sample is most thoroughly
studied in terms of stellar multiplicity in a wide range of
orbital periods (orbital axes) among the Population II field
stars.

As a result of observations of the sample stars using the method of
high angular resolution on the BTA, we detected 20 speckle
interferometric components for 19 primaries. Seven of them were known
as spectroscopic pairs and four --- as astrometric binaries (Fig.
\ref{diff_meth_sys}). Nine systems were resolved for the first
time: 5 binaries (\object{G191-55}, \object{G114-25}, \object{G142-44},
\object{G28-43}, \object{G130-7}), 3 triples (\object{G87-47}, \object{G111-38},
\object{G190-10}) and one quadruple (\object{G89-14}).

Combining different research methods allows us to estimate the
frequency of the systems of different multiplicity with the
orbital semi-major axes ranging from a few to tens of thousands
of astronomical units. The ratio of single, binary, triple and
quadruple systems among 221 primary components in our sample
amounts to 147:64:9:1. More than half of the stars in the sample
are members of binary and multiple systems:
\begin{displaymath}
f_{stars}=\frac{2B+3T+4Q}{S+2B+3T+4Q}\approx52\%\pm6\%.
\end{displaymath}
Multiplicity of the sample is
\begin{displaymath}
f_{systems}=\frac{B+T+Q}{S+B+T+Q}\approx33\%^{+6\%}_{-7\%}.
\end{displaymath}
As before, a 95\% confidence interval was used as the error of
obtained multiplicity.
For spectroscopic systems, we made an analysis of the number of
undetected components using analytical calculations, while the
similar estimates for astrometric pairs were obtained using the
Monte Carlo numerical method. For our sample the corrected for
undetected components and selection effects (eventually
only the $\mathrm{\ddot{O}}$pik effect was accounted for)
multiplicity is $f_{systems} \approx 40\%$.
\citet{dm91} give $f_{systems} \approx 60\%$ for the thin
disc G dwarfs.

For colder subdwarfs of K--M spectral classes, \citet{jao_2009}
deduced S:B:T:Q=46:12:2:2, and $f_{systems} = 26\pm6\%$ accounting
for spectroscopic, speckle interferometric and visual data.
Within errors our result coincides with the result
obtained by \citet{jao_2009}.

Seven stars in our sample are blue stragglers. We found that with
an exception of \object{G245-32}, all of them are binaries. This supports
the hypothesis of the connection of the blue stragglers
phenomenon with their binary nature.

The ratio of binary:triple:quadruple systems (B:T:Q) among the
Population II stars in our sample is 64:9:1. This can be compared
with the ratio 40:7:2 for the Population I stars \citep{dm91}.
The difference between the ratios is statistically
indistinguishable. Therefore, we came to a conclusion that a
stable hierarchical multiple system is the universal evolutional
outcome of the star formation process both at the time of the
creation of our Galaxy and nowadays.

Three of the resolved by us binary systems, \object{G76-21} (\object{HIP 12529}),
\object{G114-25} (\object{HIP 44111}) and \object{G217-8} (\object{HIP 115704}),
have very low metallicities ($\mathrm{[m/H]}<-2$). A triple system \object{G40-14}
also belongs to this metallicity range (see Table \ref{multiple}).
Altogether, 63 primaries from our sample have
$\mathrm{[m/H]}<-2$. From these, 18 primaries have one companion
and one, G40-14, has two of them. To date, only a few high
multiplicity ($N>2$) systems in very low metallicity regime are
known. Further accumulation of empirical data for these objects
will help answer the question about a possible dependence of the
orbital elements distribution of binary and multiple systems on
their metallicity.

We did not find any significant differences in the multiplicity
ratios of subdwarfs moving on prograde and retrograde ($V<-220$
km/s) galactic orbits. \citet{carney_2005a} found a decreased
ratio of strongly retrograde ($V<-300$ km/s) binaries: $10\%\pm
2\%$ against $28\%\pm 3\%$ for a prograde sample. Their
conclusion is supported by our study of 11 stars with $V<-300$
km/s: only one of them is a binary.
With the increase of the norm of spatial velocity vector
$v=\sqrt{U^2+V^2+W^2}$  the frequency of double and multiple systems in
our sample falls with a probability of at least 70\%.

We have shown that the distribution of orbital periods of the old
stars differs both in shape and in the maximum's location from
that of Population I stars (Fig. \ref{periodscomparison_pic}). Most
of the detected Population II binaries have periods between 1 and
10 yrs. The period distribution for G, K and M thin disc dwarfs
does not depend on a spectral class. In the semilog scale, it can
be approximated by a Gaussian with the maximum at $\log P_{max}
\approx 5$ dex (see Fig. 1 in \citealt{kroupa_1995}). Orbital periods
of the thin disc stars are distributed more symmetrically. Compared
to Population II stars, the maximum of the distribution is
shifted towards larger $P$ by two orders.

An important feature in the period distribution of binary and
multiple subdwarfs is the lack of short-period systems with the
periods of less than 1 day. The range of $P<10$ days is
represented in our sample just by one SB2 system --- \object{G183-9}. The
reason for that lack of old short-period pairs could be in the
dynamical evolution of the systems with short orbital periods,
which could lead to a merge of the components and to the
formation of blue stragglers (e.g., \citealt{bailyn}). Another
important peculiarity in period distribution of old metal-poor
stars is the presence of couples with $P>10^{8}$ days (Fig.
\ref{periodsbin_pic} and \ref{periodsall_pic}).
It is quite possible that such enormous periods could be an error
occuring from the formula (\ref{allen_eq}). Such low-binding energy
objects survived over many billions of years and bear important
information on the mass density distribution in our Galaxy. The
orbital periods of the subdwarfs and thin disc dwarfs lie nearly
in the same range and constitute 10 orders. It is impossible to explain
such a wide range of periods by the dynamical evolution only, as
it is formed at the earliest stages of the stellar system
formation in the nuclei of molecular clouds (e.g., \citealt{kroupa_burkert}).
Our data show that the range of possible orbital
periods of binary and multiple systems is not decreasing over
billions of years of the dynamical evolution. Only the shape of
the period distribution is changing. An interpretation of the
distinctions between the period distributions of the stars of
different populations requires further study.

The substellar mass companions (brown dwarfs and planets) stayed
beyond the scope of our consideration. Current studies (e.g.,
\citealt{fischer_valenti}) testify to the existence of a
correlation between the metallicity of stars and the presence of
orbiting planets. At the time of writing, catalogs of
stars with planets \citep{exoplanet.eu,exoplanets.org} do not contain stars
with $\mathrm{[Fe/H]}<-1$. Some researchers claim that brown
dwarfs and some very low mass stars (and possibly planets?), form
a separate population with its own multiplicity and kinematical
properties (see, e.g. \citealt{kroupa_2003}). As likely as not, the
low metallicity regime may influence the formation of this
population. It is quite possible that Population II field stars
do not contain any brown dwarfs or planets as companions at all.
Another important question is whether there exist any substellar
components in the halo and thick disc systems of high
multiplicity ($N>2$).

The ratio of binary and multiple systems among the Population II
stars, found in this study, does not contradict the hypothesis
that the chemical composition of protostellar molecular clouds
makes but an insignificant impact on the star formation process.
This indicates that the halo stars were formed as a result of
fragmentation of molecular clouds' nuclei, similarly to the way
the stars form today. However, this issue remains unclear when
we consider the substellar mass regime.

The frequency of binary and multiple metal-poor stars imposes some
restrictions on the formation of the stellar halo in our Galaxy
as well. There are two general scenarios of the formation of the
Milky Way's stellar halo (see \citealt{majewski} for more details):
\begin{itemize}
  \item Most of the halo stars were born in globular clusters or dwarf galaxies, which were then
accreted and destroyed in the gravitational potential of the Milky Way \citep{bell}.
  \item The halo stars are genetically bound with our Galaxy. The accretion of globular
clusters and dwarf galaxies observed today \citep*{ibata},
produces only a small fraction of halo stars. The kinematical
structures detected in the stellar halo \citep{helmi,bell}
are a consequence of the Galaxy's gravitational
potential inhomogeneities and a manifestation of various
resonances.
\end{itemize}
Any scenario of the stellar halo formation has to impose certain
restrictions on the rate of binary and multiple systems and on
their characteristics, i.e. on the distributions of orbital
periods, component mass ratios, eccentricities, etc.
Particularly, high percentage of binary and multiple halo field
stars indicates that the formation of stellar halo via the
destruction of globular clusters is unlikely in our Galaxy, since
the relative number of binaries in globular clusters \citep{sollima_2007}
is smaller than that of metal-poor field stars. It is
currently not clear how does the dynamical evolution of globular
clusters influence the binary frequency \citep*{ivanova,hurley,sollima_2008},
therefore the scenario of the halo field subdwarfs formation through a
dissociation of globulars cannot be fully discarded.

The question of the differences between stellar streams in terms
of binary and multiple systems requires further accumulation of
observational data. Our material does not contradict neither an
assumption of the parity of binary and multiple stars frequency
in different streams, nor the hypothesis of their dynamical
origins (e.g., \citealt{famaey_2008}).

Some of the detected speckle interferometric pairs, \object{G76-21},
\object{G63-46}, \object{G28-43}, \object{G217-8}, \object{G130-7},
\object{G102-20}, \object{BD+19$^{\circ}$ 1185A}, \object{G87-47},
with presumably short orbital periods are suitable for
monitoring for orbit calculations and mass determination of the
metal-poor stars. These studies can contribute to a calibration of
the mass-luminosity relation and to the verification of the
theories of dynamical evolution.

\acknowledgments

I am grateful to all the members of the Group
of High Angular Resolution Methods in Astronomy of the SAO RAS for
their help in securing the observations made, and personally to
Yu. Balega, A. Tokovinin and an anonymous referee
for valuable comments that led to
a notable improvement of this article. I thank D. Latham, who
provided the data on spectroscopic multiplicity of objects from the
CLLA catalog along with the corrected periods of some binary
stars, V. Dyachenko for finding the components of the stellar
streams members in the literature and A. Zyazeva for
the help with translation. I acknowledge the continuous
support provided by the Russian Foundation for Basic Research
(project no. 04-02-17563) and the Programme of Physical Sciences
of the Russian Academy of Sciences.

{\it Facilities:} \facility{BAT (Speckle interferometer)}.

\appendix

\section{Impact of Galactic structure on spatial distribution of sample stars in solar neighbourhood}

To be able to estimate the number of stars in a given volume by an
extrapolation of the calculated quantity of stars in a smaller
volume, we have to take into account the structure of the Galaxy.
A null hypothesis is that in the volume of 250 pc from the Sun,
the amount of thick disc stars is $1\ 000$ times bigger than that
within 25 pc. We assume a uniform distribution of halo stars on
the scale of 100 pc$^3$ in the solar vicinity. Let us consider
how an exponential decrease in the number density with an increase
in the distance from the galactic plane in the sample of thick
disc stars affects the null hypothesis.

Let's introduce a $k=N_{1}/N_{0}$ coefficient, where
\begin{equation}
\label{N1}
 N_{1}=\int\limits_{-230}^{270}(250^2-(z -20)^2)\times e^{-\left| z
 \right|/H}dz.
\end{equation}
The $N_{1}$ is proportional to the quantity of the thick disc
stars within 250 pc from the Sun. We assume that the Sun is
located at the distance of 20 pc above the galactic plane
\citep{humphreys_larsen}. The quantity of stars, located within
25 pc from the Sun is proportional to:
\begin{equation}
\label{N2}
  N_{0}=\int\limits_{-5}^{45}(25^2-(z - 20)^2)\times e^{-\left| z
 \right|/H}dz.
\end{equation}
A true quantity of the stars in these volumes can be obtained by
a multiplication of $N_{1}$ or $N_{0}$ on $\pi \times \rho_{0}$,
where $\rho_{0}$ is the number density of thick disc stars at the
galactic plane ($z=0$). Apparently, in case of uniform
distribution of stars in space, $k$ is equal to $1\ 000$. All
values in these formulae are expressed in parsecs. The location
above the Galactic plane is designated by $z$. The thick disc
scale height $H$ varies from 1000 to 1500 pc. For our calculations
we took $H=1048$ pc \citep{veltz}. After integrating the
right parts of (\ref{N1}) and (\ref{N2}), we obtain $k=933$.
Therefore, the deviation from the homogeneous number density
distribution of thick disc stars within 250 pc from the Sun is
approximately 7\%. The median metallicity of the stars in our
sample is $\mathrm{[m/H]}\approx-1.6$. The percentage of the
metal-weak thick disc tail stars in the range of
$-1.6<\mathrm{[Fe/H]}<-1$ may reach $60-70\%$ \citep*{morrison,beers}.
Nevertheless, while considering our sample, the Galactic structure in
the solar neighbourhood can be neglected.

\clearpage

\clearpage

\clearpage

\clearpage

\clearpage

\clearpage

\clearpage

\clearpage

\clearpage

\clearpage

\clearpage

\clearpage

\begin{table}
\centering
\caption{Properties of three stellar Populations}
\label{populations_properties}
\begin{tabular}{l|c|c|c} \hline\hline
 Value              & Thin Disc & Thick Disc & Halo \\ \hline
 $\sigma_{U}$        &     $43$    &    $67$      &  $131$ \\
 $\langle V\rangle$  &    $-15$    &   $-53$      & $-226$ \\
 $\sigma_{V}$        &     $28$    &    $51$      &  $106$ \\
 $\sigma_{W}$        &     $17$    &    $42$      &   $85$ \\
 $\langle [\mathrm{Fe/H}]\rangle$ &$-0.1$&$-0.8$&$-1.8$  \\
 $\sigma_{[\mathrm{Fe/H}]}$       & $0.2$& $0.3$& $0.5$  \\
 $\mathrm{Fraction} f$            &    $0.925$  &    $0.070$   &  $0.005$\\ \hline\hline
\multicolumn{3}{l}{According to the data from \citet{robin_2003}.}
\end{tabular}
\end{table}

\clearpage

\begin{table*}
\vspace{6mm}
\centering
\caption{Multiplicity as a function of metallicity}
\label{mult_metal_tab}
\vspace{5mm}\begin{tabular}{l|c|c|c|c} \hline\hline
$\mathrm{[m/H]}$  & $(-3,-2.5]$ & $(-2.5,-2.0]$ & $(-2.0,-1.5]$ & $(-1.5,-1.0)$ \\ \hline
$S:B:T:Q$      &  17:5:1:0   & 26:13:0:0     & 41:20:2:1     & 62:26:6:0 \\
${f_{systems}}^{\ast}$& $26\%^{+23\%}_{-16\%}$ & $33\%^{+17\%}_{-14\%}$ & $36\%^{+13\%}_{-12\%}$ & $34\%^{+11\%}_{-9\%}$    \\
Number of systems &  23      &  39           &  64           & 94        \\ \hline\hline
\multicolumn{2}{l}{$^{\ast}$ 95\% confidence interval is given}
\end{tabular}
\end{table*}

\clearpage

\begin{deluxetable}{l|c|c|c|c|c|c}
\tabletypesize{\scriptsize}
\rotate
\tablecaption{Multiplicity as a function of kinematics \label{mult_kinematics_tab}}
\tablewidth{0pt}
\startdata
\hline\hline
$U$  (km/s)    &$(-300,-200]$& $(-200,-100]$ & $(-100,0]$    & $(0,100]$ &$(100,200]$& $(200,300]$\\ \hline
$S:B:T:Q$      &  11:5:1:0   & 29:14:2:1     & 40:14:1:0     & 32:17:5:0 & 24:8:0:0  & 9:4:0:0 \\
$f_{systems}$& $35\%^{+27\%}_{-21\%}$ & $37\%^{+15\%}_{-14\%}$ & $27\%^{+14\%}_{-11\%}$ & $41\%^{+14\%}_{-13\%}$ & $25\%^{+18\%}_{-14\%}$ & $31\%^{+30\%}_{-22\%}$ \\
Number of systems &  17      &    46         &     55        &    54     &    32     & 13 \\ \hline\hline

$V$  (km/s)    &$(-300,-250]$& $(-250,-200]$ & $(-200,-150]$ &$(-150,-100]$&$(-100,-50]$& $(-50,0]$\\ \hline
$S:B:T:Q$      &  9:6:0:0    &   25:8:1:0    &   34:15:1:0   &  27:10:4:0  &  25:11:1:1 & 13:10:1:0\\
$f_{systems}$& $40\%^{+28\%}_{-24\%}$ & $26\%^{+18\%}_{-13\%}$ & $32\%^{+15\%}_{-12\%}$ & $34\%^{+17\%}_{-14\%}$ & $34\%^{+17\%}_{-14\%}$ & $46\%^{+21\%}_{-20\%}$   \\
Number of systems &   15     &    34         &      50       &     41      &     38     &   24     \\ \hline\hline

$W$  (km/s)    &$(-300,-100]$& $(-100,-50]$ & $(-50,0]$ & $(0,50]$  & $(50,100]$ & \\ \hline
$S:B:T:Q$      &  11:2:1:0   &   18:13:1:0  & 41:18:4:0 & 40:13:2:1 & 29:14:1:0  & \\
$f_{systems}$& $21\%^{+30\%}_{-16\%}$ & $44\%^{+18\%}_{-18\%}$ & $35\%^{+13\%}_{-12\%}$ & $29\%^{+13\%}_{-12\%}$ & $34\%^{+16\%}_{-14\%}$ & \\
Number of systems &   14     &     32       &     63    &   56      &     44     & \\ \hline\hline

$\sqrt{U^2+V^2+W^2}$ (km/s) & $(0,100]$& $(100,200]$ & $(200,300]$ & $(300,400]$  & $>400$ & \\ \hline
$S:B:T:Q$      &  10:8:2:0   &   48:23:5:1  & 57:24:2:0 & 26:7:0:0 & 5:0:0:0  & \\
$f_{systems}$& $50\%^{+23\%}_{-23\%}$ & $38\%^{+11\%}_{-11\%}$ & $31\%^{+11\%}_{-9\%}$ & $21\%^{+18\%}_{-12\%}$ & $0\%^{+52\%}_{-0\%}$ & \\
Number of systems &   20     &      77      &    83    &    33     &    5      & \\ \hline\hline
\enddata
\tablecomments{For $f_{systems}$ the 95\% confidence interval is given.}
\end{deluxetable}

\clearpage

\begin{table}
\vspace{6mm}
\centering
\caption{Population II multiple systems in our survey}
\label{multiple}
\vspace{5mm}\begin{tabular}{l|c|c|c|c} \hline\hline
\multirow{2}{*}{Name}  &\multirow{2}{*}{$m_{V}$} & \multirow{2}{*}{[m/H]$^{\star}$}   & Distance$^{\star}$& \multirow{2}{*}{Multiplicity} \\
                     &  &         &  (pc)  & \\ \hline
\object{G95-57}               &   8\fm78  & $-1.05$ & 25 & 3 \\
\object{BD$+19^{\circ}$ 1185} &   9.3   & $-1.47$ & 42 & 3 \\
\object{G89-14}               &   10.4  & $-1.9$  & 94 & 4 \\
\object{G87-45}               &   11.44 & $-1.49$ & 123 & 3 \\
\object{G87-47}               &   10.34 & $-1.34$ & 62 & 3 \\
\object{G111-38}              &   9.11  & $-1.04$ & 50 & 3 \\
\object{G40-14}               &   11.2  & $-2.71$ & 164 & 3 \\
\object{G59-1}                &   9.52  & $-1.14$ & 50 & 3 \\
\object{G17-25}               &   9.63  & $-1.54$ & 35 & 3 \\
\object{G190-10}              &   11.22 & $-1.92$ & 91 & 3 \\ \hline\hline
\multicolumn{5}{l}{$\star$ According to the data from CLLA.}
\end{tabular}
\end{table}

\clearpage

\begin{table*}
\vspace{6mm}
\centering
\caption{Moving groups in our sample}
\label{streams}
\vspace{5mm}\begin{tabular}{l|c|c|c|c|c|c} \hline\hline
Group          & [Fe/H]                  & U                         & V                              & W                           & $n_1:n_2:n_3$          & $N_1:N_2:N_3$           \\ \hline
Ross 451       & $-1.5$                  & $-89\pm104$               & $-346\pm5$                     & $-37\pm77$                  & 8:1:1                  & 36:5:1                  \\
Kapteyn's      & \multirow{2}{*}{$-1.5$} & \multirow{2}{*}{$13\pm82$}& \multirow{2}{*}{$-288.5\pm6.0$}& \multirow{2}{*}{$-16\pm67$} & \multirow{2}{*}{4:2:0} & \multirow{2}{*}{29:4:0} \\
star           &                         &                           &                                &                             &                        &                         \\ \hline\hline
\multicolumn{5}{l}{Data are taken from \cite{eggen_1996a,eggen_1996b}.}
\end{tabular}
\end{table*}

\end{document}